\documentclass[preprint,12pt]{elsarticle}

\newtheorem{theorem}{Theorem}
\newenvironment{proof}[1][Proof]{\textbf{#1.} }{\ \rule{0.5em}{0.5em}}
\usepackage{amssymb}
\usepackage{amsmath}
\usepackage{xfrac}
\usepackage[linesnumbered,lined,boxed,commentsnumbered,ruled,vlined]{algorithm2e}
\usepackage{subfig}
\usepackage{graphicx}

\journal{Journal of Applied Statistics}

\begin{document}

\begin{frontmatter}



\title{\bf Adaptive Bayesian Structure Learning of DAGs With Non-conjugate Prior
}


\author[inst1]{S. Nazari\footnote{Email: asemaneh1369@gmail.com}}

\affiliation[inst1]{organization={Department of Statistics, Faculty of Mathematical Sciences, Ferdowsi University of Mashhad, Iran}},

\author[inst1]{M. Arashi\footnote{Corresponding Author, Email: arashi@um.ac.ir}}

\author[inst2]{A. Sadeghkhani\footnote{Email: asadeghkhani@ncat.edu}}

\affiliation[inst2]{organization={Department of Mathematics and Statistics, North Carolina Agricultural and Technical State University, USA}}

\begin{abstract}
\noindent {\it Abstract:}
Directed Acyclic Graphs (DAGs) are solid structures that describe and infer variables' dependencies in multivariate scenarios. A thorough comprehension of the accurate DAG-generating model is crucial for causal discovery and estimation. Our work suggests utilizing a non-conjugate prior for Gaussian DAG structure learning to enhance the posterior probability. We employ the idea of using the Bessel function to address the computational burden, providing faster MCMC computation than conjugate priors. In addition, our proposal exhibits a greater rate of adaptation when compared to the conjugate prior, specifically for including nodes in the DAG-generating model. Simulation studies demonstrate the superior accuracy of DAG learning, and we obtain the same maximum a posteriori and median probability model estimate for the AML data using the non-conjugate prior. 
\end{abstract}

\begin{graphicalabstract}
\includegraphics[scale=0.99]{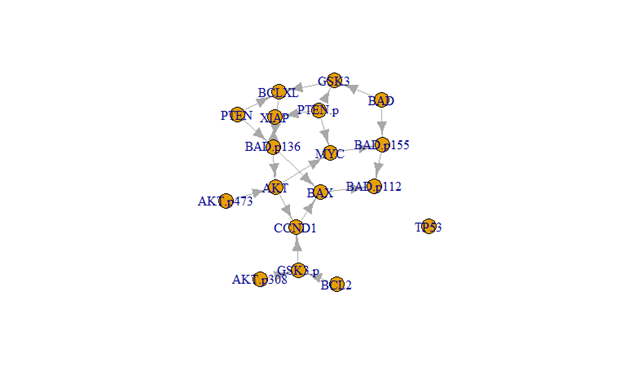}
\end{graphicalabstract}

\begin{highlights}
\item We develop a Bayesian structure learning of DAGs with high posterior probability.
\item We combat computational burden of the Gaussian DAG learning with non-conjugate priors.
\item The proposed Bayesian structure learning of Gaussian DAGs include nodes if the inclusion probability is high, unlike the case of using conjugate priors.
\end{highlights}

\begin{keyword}
Bayesian structure learning \sep Directed acyclic graph \sep Graphical model \sep Normal-gamma prior \sep Posterior probability.
\end{keyword}

\end{frontmatter}


\section{Introduction}
\label{sec:sample1}
In recent decades, probabilistic graphical models have emerged as potent and valuable instruments for representing and deducing interdependent relationships in complex multivariate settings (\cite{Lauritzen}). Directed acyclic graphs (DAGs), called Bayesian networks, are commonly used to determine a specific set of conditional independence relationships between variables while adhering to the DAG Markov property. A directed acyclic graph (DAG) establishes conditional independencies across variables by factorizing the joint distribution based on the DAG structure. It is useful for identifying these relationships straight from the graph using specific criteria.
The empirical evidence suggests that the set of conditional dependencies among variables cannot be predetermined in real-world data situations, thereby rendering the DAG-generating model uncertain.
The primary goal of analysis in various domains is to perform DAG structural learning, which involves identifying dependent links among variables. This process is crucial for gaining insights into the behavior of these variables. For further information, refer to the works of \cite{Friedman} and \cite{Sachs}.

Moreover, integrating DAGs with the do-calculus theory (\cite{Pearl}) enables the utilization of DAGs for causal inference, which aids in detecting and estimating causal links between variables. Learning the graphical structure of DAGs when causal assumptions are given is widely known as causal structure learning or causal discovery (\cite{Maatuis} and \cite{Peters}). Based on the concepts of fidelity and causal sufficiency (\cite{Spirtes}), the structure of a DAG can only be determined up to its Markov equivalence class (\cite{Andersson}). This class includes all DAGs that have the same Markov property. Nevertheless, as stated in \cite{Pearl}, it is important to note that even though comparable DAGs may represent various Structural Causal Models (SCMs), the Markov equivalence class that has been established can still be utilized to uncover a set of possibly unique causal effects. 

From a Bayesian standpoint, learning the structure of a DAG is typically approached as a form of Bayesian model selection. Within this approach, learning the structure of DAGs involves estimating a posterior distribution throughout the range of possible graphs. This method systematically measures the uncertainty surrounding the model that generates the data. Markov Chain Monte Carlo (MCMC) methods are frequently used to estimate the posterior distribution of graph structures, as demonstrated in research such as \cite{Cooper} \cite{Ni},  \cite{Castelletti}, \cite{Castelletti1}, and \cite{Castelletti2}.

On the other hand, Bayesian approaches necessitate collecting prior information for each model-specific parameter, which must adhere to the limitations given by the conditional independencies of the underlying DAG structure, as discussed in Geiger's work (\cite{Geiger}).

From this viewpoint, the specification of the prior distribution plays a crucial role. The paper \cite{Castelletti1} uses a normal-inverse-gamma conjugate prior to learning the structure of Gaussian DAGs. Conjugate priors generally provide manageable analytical calculations, making MCMC computations easier. In this study, we utilize a non-conjugate normal-gamma prior, as described by \cite{Bekker}, to enhance the posterior probability and the graphical matrices in learning the structure of Gaussian DAGs. We prove that our method displays cautious behavior regarding included edges in the Gaussian DAG. Please refer to Table 3 and the ensuing discussion for more details.
It offers expedited computation in comparison to the normal-inverse-gamma model.
In addition, \cite{Mohammadi}, \cite{Mohammadi1}, and \cite{Vogels} examine the process of Bayesian structure learning in undirected Gaussian graphical models.

The organization of the rest of this study is as follows. In section 2, we give some preliminary results on graphical models and DAGs building on the Bayesian structure studied in \cite{Castelletti1}. Section 3 gives posterior probabilities and marginal distributions for the normal-gamma prior on the Gaussian DAG parameters. An MCMC algorithm is given in section 4, while Section 5 contains numerical analyses. We conclude our study in section 6, with the proof of main result is provided in Appendix.
\setcounter{equation}{0}
\section{Preliminaries}
In this section, we present fundamental concepts of DAGs and graph theory. Readers are referred to \cite{Lauritzen} and \cite{Pearl1} for details. 
\subsection{Gaussian DAG-models}
Consider 
$\mathcal{D}=\left(V,E\right)$ be a Directed Acyclic Graph (DAG), with 
$V=\lbrace 1,\ldots,q \rbrace$ 
representing a collection of vertices (or nodes) and
$E\subseteq V\times V$
is a set of edges. If 
$\left(u,v\right)\in E$, 
$\mathcal{D}$
contains the directed edge 
$u\rightarrow v$. 
Furthermore, 
$\mathcal{D}$
cannot have cycles, that is paths of form 
$u_{0}\rightarrow u_{1}\rightarrow \ldots \rightarrow u_{k}$
where 
$u_{0}\equiv u_{k}$. For a given DAG 
$\mathcal{D}$, 
if there is an edge
$u\rightarrow v$ we say that $u$ is a $parent$ of $v$ (conversely, $v$ is a child of $u$) and denote the parent set of $v$ in 
$\mathcal{D}$  as $pa_{\mathcal{D}}\left( v\right)$, while the set $fa_{\mathcal{D}}\left( v\right)=v\cup pa_{\mathcal{D}}\left( v\right)$ is called the $family$ of $v$ in $\mathcal{D}$. Furthermore, a DAG is $complete$ if all its nodes are joined by edges. Finally, a DAG
$\mathcal{D}$
can be uniquely expressed by its 
$\left(q,q\right)$\textit{ adjacency matrix}
$\boldsymbol{A}^{\mathcal{D}}$
such that 
$\boldsymbol{A}_{u,v}^{\mathcal{D}}=1$
if and only $\mathcal{D}$ contains $u\rightarrow v$ and 0 otherwise.\\
A DAG $\mathcal{D}$ encodes a set of conditional independencies between nodes(variables) that can be read-off from the DAG using graphical criteria, such as $d$-$separation$ \cite{Pearl}. The DAG Markov property is defined by the resulting set of conditional independence encoded in
$\mathcal{D}$.\\
In the following, we define a Gaussian DAG-model in terms of likelihood and prior
distributions for model parameters.\\
Let 
$\mathcal{D}=\left(V,E\right)$ be a DAG, 
$\boldsymbol{x}=\left(x_{1},\dots,x_{q}\right)^{T}$ a collection of real-valued random variables each associated to a node in V. Also assume that the joint density of 
$\left(x_{1},\dots,x_{q}\right)^{T}$
belongs to a zero-mean Gaussian DAG-model, namely
\begin{equation}\label{eq1}
	\boldsymbol{x}\mid\boldsymbol{\Omega}_{\mathcal{D}}\sim\mathcal{N}_{q}\left(0,\boldsymbol{\Omega}_{\mathcal{D}}^{-1}\right),\qquad\boldsymbol{\Omega}_{\mathcal{D}}\in\mathcal{P}_{\mathcal{D}},
\end{equation}
where $\boldsymbol{\Omega}_{\mathcal{D}}=\boldsymbol{\Sigma}_{\mathcal{D}}^{-1}$
is the precision (inverse-covariance) matrix, and 
$\mathcal{P}_{\mathcal{D}}$ is the space of symmetric positive definite (s.p.d) precision matrices Markov w.r.t. 
$\mathcal{D}$. Therefore, $\boldsymbol{\Omega}_{\mathcal{D}}$ satisfies the conditional independencies (Markov property) embedded by 
$\mathcal{D}$.\\
According to modified Cholesky decomposition of 
$\boldsymbol{\Omega}=\boldsymbol{\Sigma}^{-1}$, that is
$\boldsymbol{\Omega}=\boldsymbol{L}\boldsymbol{D}^{-1}\boldsymbol{L}^{T}$, 
model \eqref{eq1} can be written as follows:
\begin{equation}\label{eq2}
	\boldsymbol{L}^{T}\left(x_{1},\dots,x_{q}\right)^{T}=\boldsymbol{\varepsilon},\qquad\boldsymbol{\varepsilon}\sim\mathcal{N}_{q}\left(\boldsymbol{0},\boldsymbol{D}\right),
\end{equation}
where
$\boldsymbol{L}$
is a 
$q\times q$ matrix of coefficients that has a value for every 
$\left(u,v\right)$-element 
$\boldsymbol{L}_{uv}$ with $u\neq v$, 
$\boldsymbol{L}_{uv}\neq 0$ if and only if 
$\left(u,v\right)\in E$, whereas 
$\boldsymbol{L}_{uu}=1$ for each 
$u=1,\dots q$. Also $\boldsymbol{D}$
is a $q\times q$ diagonal matrix with 
$\left(u,u\right)$-element 
$\boldsymbol{D}_{uu}$.\\
As a result, it deduces a re-parameterization of 
$\boldsymbol{\Omega}=\boldsymbol{\Sigma}^{-1}$
in relation to the node-parameters 
$\lbrace\left(\boldsymbol{D}_{jj},\boldsymbol{L}_{\prec j\rbrack}\right),j=1,\dots,q\rbrace$, such that 
\begin{equation*}
	\boldsymbol{L}_{\prec j]}=-\boldsymbol{\Sigma}_{\prec j\succ}^{-1}\boldsymbol{\Sigma}_{\prec j]},\qquad \boldsymbol{D}_{jj}=\boldsymbol{\Sigma}_{jj|pa_{\mathcal{D}}\left(j\right)}, 
\end{equation*}
Where 
$\boldsymbol{\Sigma}_{jj|pa_{\mathcal{D}}\left(j\right)}=\boldsymbol{\Sigma}_{jj}-\boldsymbol{\Sigma}_{[ j\succ}\boldsymbol{\Sigma}_{\prec j\succ}^{-1}\boldsymbol{\Sigma}_{\prec j]},{\prec j]}=pa_{\mathcal{D}}\left(j\right)\times j,\,[j\succ=j\times pa_{\mathcal{D}}\left(j\right),\,\prec j \succ = pa_{\mathcal{D}}\left(j\right)\times pa_{\mathcal{D}}\left(j\right)$. As well parameter  
$\boldsymbol{D}_{jj}$ represents the conditional variance of 
$x_{j}$, $\mathbb{V}ar\left(x_{j}\mid\boldsymbol{x}_{pa_{\mathcal{D}}\left(j\right)}\right)$.
For further details refer to \cite{Castelletti1}.\\
Model \eqref{eq1} can be re-written as follows based on \eqref{eq2}: 
\begin{equation}\label{eq3}
	f\left(x_{1},\ldots,x_{q}\mid\boldsymbol{D},\boldsymbol{L}\right)=\prod_{j=1}^{q}d\mathcal{N}\left(x_{j}|-\boldsymbol{L}_{\prec  j]}^{T}\boldsymbol{x}_{pa_{\mathcal{D}}\left(j\right)},\boldsymbol{D}_{jj}\right).
\end{equation}
Now consider $n$ random samples 
$\boldsymbol{x}_{i}=\left(x_{i,1},\ldots,x_{i,q}\right)^{T}\in\mathbb{R}^{q},\, {i}=1,\ldots,n$, from (\ref{eq3}), and suppose 
$\boldsymbol{X}=\left(\boldsymbol{x}_{1},\dots,\boldsymbol{x}_{n}\right)^{T}$ be the 
$n\times q$ data matrix. The likelihood function is given by 
\begin{align}\label{eq4}
	f\left(\boldsymbol{X}\vert\boldsymbol{D},\boldsymbol{L}\right)&=\prod_{i=1}^{n}f\left(x_{i,1},\ldots,x_{i,q}\mid\boldsymbol{D},\boldsymbol{L}\right)\nonumber\\ &=\prod_{j=1}^{q}d\mathcal{N}_{n}\left(\boldsymbol{X}_{j}\mid-\boldsymbol{X}_{pa_{\mathcal{D}}\left(j\right)}\boldsymbol{L}_{\prec j]},\boldsymbol{D}_{jj}\boldsymbol{I}_{n}\right),
\end{align}
where 
$\boldsymbol{X}_{A}$
is the $n\times\lvert A \rvert$ sub-matrix of $\boldsymbol{X}$ corresponding to the set $A$ of columns of 
$\boldsymbol{X}$
and $\boldsymbol{I}_{n}$
denotes the $n\times n$
identity matrix.\\
In a recent study, \cite{Castelletti1} introduced a conjugate prior for node-parameters 
$\lbrace\left(\boldsymbol{D}_{jj},\boldsymbol{L}_{\prec j ]}\right),\,j=1,\ldots,q\rbrace$,
that are priori independent with distribution 
\begin{align}\label{eq5}
	\boldsymbol{D}_{jj}\mid \mathcal{D}&\sim \Gamma^{-1}\left(\frac{1}{2}\alpha_{j}^{\mathcal{D}},\frac{1}{2}\boldsymbol{U}_{jj|pa_{\mathcal{D}}\left(j\right)}\right),\nonumber\\ 
	\boldsymbol{L}_{\prec j ]}\mid\boldsymbol{D}_{jj},\mathcal{D}&\sim\mathcal{N}_{\lvert pa_{\mathcal{D}}\left(j\right)\rvert}\left(-\boldsymbol{U}_{\prec j \succ}^{-1}\boldsymbol{U}_{\prec j ]},\boldsymbol{D}_{jj}\boldsymbol{U}_{\prec j \succ}^{-1}\right),
\end{align}
where $\boldsymbol{U}$ is the rate hyperparameter (a $q\times q$ symmetric positive definite matrix) and $\boldsymbol{\alpha}^{\mathcal{D}}=\left(\alpha_{1}^{\mathcal{D}},\ldots,\alpha_{q}^{\mathcal{D}}\right)^{T}$ is the shape hyperparameter according to \cite{Ben}.\\
The default selection, 
$\alpha_{j}^{\mathcal{D}}=\alpha+\lvert pa_{\mathcal{D}}\left(j\right)\rvert-q+1,\,\left(\alpha>q-1\right)$
ensures compatibility amongst prior distributions for Markov equivalent DAGs. 
It can be demonstrated, in specific, that with this choice, any two Markov equivalent DAGs are allocated the same marginal likelihood.
\section{Normal-Gamma Prior}
In this section, we change the distribution of diagonal elements  $\boldsymbol{D}_{jj}=\boldsymbol{\Sigma}_{jj|pa_{\mathcal{D}}\left(j\right)}$ and compute the marginal and posterior density functions. Specifically, we replace the inverse gamma with the gamma distribution. 

Assume 
\begin{align}\label{eq7}
	\boldsymbol{D}_{jj}\mid \mathcal{D}&\sim \Gamma\left(\frac{1}{2}\alpha_{j}^{\mathcal{D}},\frac{1}{2}\boldsymbol{U}_{jj|pa_{\mathcal{D}}\left(j\right)}\right),\nonumber\\ 
	\boldsymbol{L}_{\prec j ]}\mid\boldsymbol{D}_{jj},\mathcal{D}&\sim\mathcal{N}_{\lvert pa_{\mathcal{D}}\left(j\right)\rvert}\left(-\boldsymbol{U}_{\prec j \succ}^{-1}\boldsymbol{U}_{\prec j ]},\boldsymbol{D}_{jj}\boldsymbol{U}_{\prec j \succ}^{-1}\right),
\end{align}
is a prior on parameters 
$\lbrace\left(\boldsymbol{D}_{jj},\boldsymbol{L}_{\prec j ]}\right),\,j=1,\ldots,q\rbrace$.\\
Further, assume the prior on 
$\left(\boldsymbol{D},\boldsymbol{L}\right)$
is given by 
\begin{equation}\label{eq9}
	p\left(\boldsymbol{D},\boldsymbol{L}\mid\mathcal{D}\right)=\prod_{j=1}^{q}p\left(\boldsymbol{L}_{\prec j ]}\mid\boldsymbol{D}_{jj}\right)p\left(\boldsymbol{D}_{jj}\right).
\end{equation}
In the following, we present the posterior distribution of node-parameters and marginal likelihood of each node in DAG $\mathcal{D}$.
\begin{theorem}\label{th2}
	Assuming equations \eqref{eq4} and \eqref{eq9}, the node-marginal likelihood is given by
\begin{eqnarray*}
m\left(\boldsymbol{X}_{j}\mid\boldsymbol{X}_{pa_{\mathcal{D}}\left(j\right)},\mathcal{D}\right)&=&\left(2\pi\right)^{-\tfrac{n}{2}}\times\dfrac{\big\lvert \boldsymbol{U}_{\prec j \succ} \big\rvert^{\tfrac{1}{2}}}{\big\lvert \widetilde{\boldsymbol{U}}_{\prec j \succ} \big\rvert^{\tfrac{1}{2}}}\times\dfrac{\bigg(\dfrac{1}{2}\boldsymbol{U}_{jj\mid pa_{\mathcal{D}}\left(j\right)}\bigg)^{\tfrac{1}{2}\alpha_{j}^{\mathcal{D}}}}{\Gamma\left(\tfrac{1}{2}\alpha_{j}^{\mathcal{D}}\right)}\cr
&&\times 2K_{a}\bigg(2\sqrt{\alpha\beta}\bigg)\bigg(\dfrac{\alpha}{\beta}\bigg)^{\tfrac{a}{2}}.
\end{eqnarray*}
\end{theorem}
See the Appendix for the proof.
\begin{theorem}\label{th1}
	Under the assumptions of section 2, assume \eqref{eq9} for the prior on 
	$\left(\boldsymbol{D},\boldsymbol{L}\right)$. Then the joint posterior density function of 
	$\left(\boldsymbol{D},\boldsymbol{L}\right)$
	given the data
	$\boldsymbol{X}$,
	$p\left(\boldsymbol{D},\boldsymbol{L}\mid\mathcal{D},\boldsymbol{X}\right)$,
	for 
	$j=1,\dots,q$
	is
	\begin{multline*}
		p\left(\boldsymbol{L}_{\prec j ]}\mid\boldsymbol{D}_{jj},\mathcal{D},\boldsymbol{X}\right)=\\
		\hspace{1.5cm}\left(2\pi\right)^{-\tfrac{\lvert pa_{\mathcal{D}}\left(j\right) \rvert}{2}}\boldsymbol{D}_{jj}^{-\tfrac{\lvert pa_{\mathcal{D}}\left(j\right) \rvert}{2}}\lvert \widetilde{\boldsymbol{U}}_{\prec j \succ} \rvert^{\tfrac{1}{2}}
		\exp\biggl\lbrace{-\dfrac{1}{2\boldsymbol{D}_{jj}}\left(\boldsymbol{L}_{\prec j ]}-\boldsymbol{M}^{-1}\boldsymbol{b}\right)^{T}\boldsymbol{M}\left(\boldsymbol{L}_{\prec j ]}-\boldsymbol{M}^{-1}\boldsymbol{b}\right)}\biggr\rbrace,
	\end{multline*}
	\begin{multline*}
		p\left(\boldsymbol{D}_{jj}\mid,\mathcal{D},\boldsymbol{X}\right)=\\
		\hspace{1.5cm}\dfrac{\boldsymbol{D}_{jj}^{\tfrac{1}{2}\alpha_{j}^{\mathcal{D}}-\tfrac{n}{2}-1}\exp\biggl\lbrace -\dfrac{1}{2}\boldsymbol{U}_{jj\mid pa_{\mathcal{D}}\left(j\right)}\boldsymbol{D}_{jj}-\dfrac{1}{2\boldsymbol{D}_{jj}}\biggl\lbrack -\boldsymbol{b}^{T}\boldsymbol{M}^{-1}\boldsymbol{b}+{\boldsymbol{U}}_{\prec j ]}^{T}{\boldsymbol{U}}_{\prec j \succ}^{-1}{\boldsymbol{U}}_{\prec j ]}+\boldsymbol{X}_{j}^{T}\boldsymbol{I}_{n}\boldsymbol{X}_{j}\biggr\rbrack\biggr\rbrace}{2K_{a}\left(2\sqrt{\alpha\beta}\right)\left(\dfrac{\alpha}{\beta}\right)^{\tfrac{a}{2}}},
	\end{multline*}	
	where 
	\begin{align*}
		\boldsymbol{M}&=\widetilde{\boldsymbol{U}}_{\prec j \succ}=\boldsymbol{U}_{\prec j \succ}+\boldsymbol{X}_{pa_{\mathcal{D}}\left(j\right)}^{T}\boldsymbol{I}_{n}\boldsymbol{X}_{pa_{\mathcal{D}}\left(j\right)},\,
		\boldsymbol{b}=-\boldsymbol{X}_{pa_{\mathcal{D}}\left(j\right)}^{T}\boldsymbol{I}_{n}\boldsymbol{X}_{j}-\boldsymbol{U}_{\prec j ]},\,a=\tfrac{1}{2}\alpha_{j}^{\mathcal{D}}-\tfrac{n}{2},\\
		&\alpha=\dfrac{1}{2}\biggl\lbrack -\boldsymbol{b}^{T}\boldsymbol{M}^{-1}\boldsymbol{b}+{\boldsymbol{U}}_{\prec j ]}^{T}{\boldsymbol{U}}_{\prec j \succ}^{-1}{\boldsymbol{U}}_{\prec j ]}+\boldsymbol{X}_{j}^{T}\boldsymbol{I}_{n}\boldsymbol{X}_{j}\biggr\rbrack,\,\beta=\dfrac{1}{2}\boldsymbol{U}_{jj\mid pa_{\mathcal{D}}\left(j\right)},
	\end{align*}
	and $K_{\nu}\left(.\right)$ is the Bessel function of the third kind.
\end{theorem}
The proof directly follows from  Theorem \ref{th2} and is omitted to save space.
\section{MCMC Scheme}
In this section, we detail the MCMC scheme to generate samples from the posterior distribution 
\begin{equation}\label{eq12}
	p\left(\boldsymbol{D},\boldsymbol{L},\mathcal{D}\mid\boldsymbol{X}\right)\propto f\left(\boldsymbol{X}\mid\boldsymbol{D},\boldsymbol{L},\mathcal{D}\right)p\left(\boldsymbol{D},\boldsymbol{L}\mid\mathcal{D}\right)p\left(\mathcal{D}\right),
\end{equation}
where, following \cite{Castelletti1} to each DAG $\mathcal{D}\in\mathcal{S}_{q}$, the set of all DAGs on $q$ nodes, we assume 
$p\left(\mathcal{D}\right)\propto p\left(\boldsymbol{S}^{\mathcal{D}}\right)$ with
\begin{equation}\label{eq15}
	p\left(\boldsymbol{S}^{\mathcal{D}}\right)=\pi^{\lvert\boldsymbol{S}^{\mathcal{D}}\rvert}\left(1-\pi\right)^{\tfrac{q\left(q-1\right)}{2}-\lvert\boldsymbol{S}^{\mathcal{D}}\rvert},
\end{equation}
where $\boldsymbol{S}^{\mathcal{D}}$ is the (symmetric) $0-1$ adjacency matrix of the skeleton of the DAG, $\lvert\boldsymbol{S}^{\mathcal{D}}\rvert$ is the number of edges
in the skeleton, $\pi\in\lbrack0,1\rbrack$ prior probability of edge inclusion and $\sfrac{q\left(q-1\right)}{2}$ corresponds to the maximum number of edges in a DAG on $q$ nodes.\\
In the first step, at each iteration of the MCMC scheme, we start defining a new DAG $\mathcal{D}^{'}$ from a suitable proposal distribution $q\left(\mathcal{D}^{'}\mid\mathcal{D}\right)$ on the DAG space, determining the transitions between DAGs within the space $\mathcal{S}_{q}$.\\
To do this, we examine three types of operators that locally modify an input DAG $\mathcal{D}$. These operators include inserting a directed edge (InsertD $u\rightarrow v$ for short), deleting a directed edge (DeleteD $u\rightarrow v$), and reversing a directed edge (ReverseD $u\rightarrow v$). For each $\mathcal{D}\in\mathcal{S}_{q}$, we then construct the set of \textit{valid} operators $\mathcal{O}_{\mathcal{D}}$, that is operators whose resulting graph is a DAG. Therefore, given the current $\mathcal{D}$ we suggest $\mathcal{D}^{'}$ by uniformly sampling a DAG in $\mathcal{O}_{\mathcal{D}}$. Due to one-one correspondence between each operator and resulting DAG $\mathcal{D}^{'}$, the probability of transition is given by 
$q\left(\mathcal{D}^{'}\mid\mathcal{D}\right)=\sfrac{1}{\lvert\mathcal{O}_{\mathcal{D}}\rvert}$.\\
\begin{algorithm}[H]\label{Algo1}\small
	\SetAlgoLined
	\SetKwData{Left}{left}\SetKwData{This}{this}\SetKwData{Up}{up}
	\SetKwFunction{Union}{Union}\SetKwFunction{FindCompress}{FindCompress}
	\SetKwInOut{Input}{input}\SetKwInOut{Output}{output}
	\Input{A DAG $\mathcal{D}$}
	\Output{The collection of valid operators $\mathcal{O}_{\mathcal{D}}$, a DAG $\mathcal{D}^{'}$ taken from $q\left(\mathcal{D}^{'}\mid\mathcal{D}\right)$}
	Construct:\\
	$I_{\mathcal{D}}$, the set of all possible operators of type InsertD,\\
	$E_{\mathcal{D}}$, the set of all possible operators of type DeleteD,\\
	$R_{\mathcal{D}}$, the set of all possible operators of type ReverseD;\\
	\For{each operator $o_{D}\in I_{\mathcal{D}}$}{
		add $o_{D}$ to $\mathcal{O}_{\mathcal{D}}$ if $o_{D}$ is valid
	}
	\For{each operator $o_{D}\in E_{\mathcal{D}}$}{
		add $o_{D}$ to $\mathcal{O}_{\mathcal{D}}$ if $o_{D}$ is valid
	}
	\For{each operator $o_{D}\in R_{\mathcal{D}}$}{
		add $o_{D}$ to $\mathcal{O}_{\mathcal{D}}$ if $o_{D}$ is valid
	}
	Draw uniformly an operator $o_{D}$ from $\mathcal{O}_{\mathcal{D}}$ and obtain $\mathcal{D}^{'}$ by applying it to $\mathcal{D}$
	\caption{Production of $\mathcal{O}_{\mathcal{D}}$ and sampling of $\mathcal{D}^{'}$ from $q\left(\mathcal{D}^{'}\mid\mathcal{D}\right)$}
\end{algorithm}
\vspace{0.5cm}
On the other hand, to update the DAGs, due to the structure of proposal distribution, when proposing a DAG $\mathcal{D}^{'}$ which differs from the current graph $\mathcal{D}$ by one edge $\left(h,j\right)$, we will need to compare two DAGs $\mathcal{D}$ and $\mathcal{D}^{'}$. Therefore, using Theorem \ref{th2}, $q\left(\mathcal{D}^{'}\mid\mathcal{D}\right)=\sfrac{1}{\lvert\mathcal{O}_{\mathcal{D}}\rvert}$, and \eqref{eq15} the acceptance probability for $\mathcal{D}^{'}$ is given by 
\begin{equation}\label{eq13}
	\alpha_{\mathcal{D}^{'}}=\min\biggl\lbrace 1, \dfrac{m\left(\boldsymbol{X}_{j}\mid\boldsymbol{X}_{pa_{\mathcal{D}^{'}}\left(j\right)},\mathcal{D}^{'}\right)}{m\left(\boldsymbol{X}_{j}\mid\boldsymbol{X}_{pa_{\mathcal{D}}\left(j\right)},\mathcal{D}\right)}.\dfrac{p\left(\mathcal{D}^{'}\right)}{p\left(\mathcal{D}\right)}.\dfrac{q\left(\mathcal{D}\mid\mathcal{D}^{'}\right)}{q\left(\mathcal{D}^{'}\mid\mathcal{D}\right)}\biggl\rbrace.
\end{equation}
Algorithm 2 summarizes our MCMC scheme for posterior inference on DAGs and DAG-parameters. In this context, let $S$, $B$, $\boldsymbol{X}$, $\alpha$,  $\boldsymbol{U}$ and $w$ denote respectively, the number of final MCMC draws from the posterior, the burn-in period, the $n\times q$ data matrix, the hyperparameters of the prior distribution and the prior probability of edge inclusion.\\
\begin{algorithm}[H]\small
	\SetAlgoLined
	\SetKwData{Left}{left}\SetKwData{This}{this}\SetKwData{Up}{up}
	\SetKwFunction{Union}{Union}\SetKwFunction{FindCompress}{FindCompress}
	\SetKwInOut{Input}{input}\SetKwInOut{Output}{output}
	\Input{$S$, $B$, $\boldsymbol{X}$, $\alpha$, $\boldsymbol{U}$, $w$}
	\Output{$S$ samples from the posterior distribution \eqref{eq12}}
	Initialize $\mathcal{D}^{0}$, e.g. the empty DAG;\\
	\For{$s=1,\dots,B+S$}{
		Sample $\mathcal{D}^{'}$ from $q\left(\mathcal{D}^{'}\mid\mathcal{D}^{\left(s-1\right)}\right)$ using Algorithm \ref{Algo1};\\
		Set $\mathcal{D}^{\left(s\right)}=\mathcal{D}^{'}$ with probability \eqref{eq13}, otherwise $\mathcal{D}^{\left(s\right)}=\mathcal{D}^{\left(s-1\right)}$;\\
		Sample $\left(\boldsymbol{D}^{\left(s\right)},\boldsymbol{L}^{\left(s\right)}\right)$ from its full conditional $p\left(\boldsymbol{D}^{\left(s\right)},\boldsymbol{L}^{\left(s\right)}\mid\mathcal{D}^{\left(s\right)},\boldsymbol{X}\right)=\prod_{j=1}^{q}p\left(\boldsymbol{L}_{\prec j ]}^{\left(s\right)}\mid\boldsymbol{D}_{jj}^{\left(s\right)},\mathcal{D}^{\left(s\right)},\boldsymbol{X}\right)p\left(\boldsymbol{D}_{jj}^{\left(s\right)}\mid\mathcal{D}^{\left(s\right)},\boldsymbol{X}\right)$;
	}
	\textbf{return} $\big\lbrace\left(\boldsymbol{D}^{\left(s\right)},\boldsymbol{L}^{\left(s\right)},\mathcal{D}^{\left(s\right)}\right), s=B+1,\dots,B+S\big\rbrace$ 
	\caption{MCMC scheme to sample from the posterior of DAGs and parameter}
\end{algorithm}
\section{Numerical Analysis} 
In this section, we compare our methodology with \cite{Castelletti1}, which we henceforth refer to as the Normal-Gamma (NG) model for easier reference, alongside the Normal-Inverse-Gamma (NIG) model.
\subsection{Simulation study}
The performance of our method is assessed through simulated data, with various scenarios considered by varying the sample size 
$n\in\lbrace200,300\rbrace$ and the number of nodes 
$q\in\lbrace40,50\rbrace$. We generate 40 DAGs using a probability of edge inclusion equal to
$\sfrac{3}{2q-2}$ to induce sparsity; see \cite{Peters1}. Then, for each DAG $\mathcal{D}$, we continue as follows.
We fix 
$\boldsymbol{D}^{\mathcal{D}}=\boldsymbol{I}_{q}$ and then following \cite{Castelletti3}, we randomly draw the non-zero elements of $\boldsymbol{L}^{\mathcal{D}}$ in the interval 
$\lbrack-2,-1\rbrack$$\cup$$\lbrack1,2\rbrack$. We finally construct the precision matrix 
$\boldsymbol{\Omega}=\boldsymbol{L}\boldsymbol{D}^{-1}\boldsymbol{L}^{T}$ and generate $n$ random samples from the Gaussian DAG-model
$\mathcal{N}_{q}\left(0,\boldsymbol{\Omega}_{\mathcal{D}}^{-1}\right)$. \\
The performance in learning the graphical structure of the DAG is evaluated by computing the misspecification rate (MISR), specificity (SP), sensitivity (SE), false negative rate (FNR), accuracy (AC) and the F1-score (F1) as defined in Table 1. 
The number pf true positive (TP), true negative (TN), false positive (FP), and false negative (FN) are denoted accordingly. 
Better classification performance is implied by values of specificity, sensitivity, accuracy and F1-score closer to one. The closer the values of the false negative rate and misspecification rate are to zero, the better.
\begin{table}[!h]\label{table1}
	\centering
	\renewcommand{\arraystretch}{1.5}
	\normalsize
		\begin{tabular}{|c|c|}
			\hline Measure & Performance function \\
			\hline SP & $\frac{\mathrm{TN}}{\mathrm{TN}+\mathrm{FP}}$\\
			\hline SE & $\frac{\mathrm{TP}}{\mathrm{TP}+\mathrm{FN}}$\\
			\hline FNR & $\frac{FP}{F P+TN}$\\[0.5em]
			\hline F1 & $\frac{TP}{TP+\sfrac{1}{2}(\mathrm{FP}+FN)}$\\
			\hline MISR & $\frac{\mathrm{FP}+ \mathrm{FN}}{\mathrm{q}\left(\mathrm{q-1}\right)}$\\
			\hline AC & $\frac{\mathrm{TP}}{\mathrm{TP}+\mathrm{FN}}$\\
			\hline
		\end{tabular}
	\caption{\normalsize Performance measures used to compare NIG and NG prior models in DAG structure learning.}
\end{table}
\begin{table}[!h]\label{table2}
	\normalsize
	\begin{center}
		\begin{tabular}{|c|c|c|c|c|}
			\hline & \multicolumn{2}{|c|}{ n=200 } & \multicolumn{2}{c|}{ n=300 } \\
			\hline & NIG & NG & NIG & NG \\
			\hline \multicolumn{5}{|c|}{$q=40$} \\
			\hline SP & 0.9811 & \textbf{0.998} & 0.978 & \textbf{0.994} \\
			\hline SE & 0.608 & 0 & 0.577 & 0.001 \\
			\hline FNR & 0.0186 & \textbf{0.001} & 0.021 & \textbf{0.005} \\
			\hline F1 & 0.476 & 0 & 0.441 & 0.002 \\
			\hline MISR & 0.0262 & \textbf{0.020} & 0.029 & \textbf{0.024} \\
			\hline AC & 0.9744 & \textbf{0.980} & 0.970 & \textbf{0.975} \\
			\hline \multicolumn{5}{|c|}{$q=50$} \\
			\hline SP & 0.9813 & \textbf{0.999} & 0.980 & \textbf{0.996} \\
			\hline SE & 0.527 & 0.0005 & 0.541 & 0.0007 \\
			\hline FNR & 0.0188 & \textbf{0.0006} & 0.019 & \textbf{0.003} \\
			\hline F1 & 0.392 & 0.041 & 0.389 & 0.044 \\
			\hline MISR & 0.0264 & \textbf{0.016} & 0.026 & \textbf{0.018} \\
			\hline AC & 0.974 & \textbf{0.984} & 0.974 & \textbf{0.981} \\
			\hline
		\end{tabular}
	\end{center}
	\caption{Performance measure for the simulated data for NIG and NG with number of nodes $q\in\lbrace 40,50\rbrace$ and sample size $n\in\lbrace 200,300\rbrace$.}
\end{table}
\begin{figure}[!ht]\hspace{-1.8cm}
	\begin{tabular}{cc}
		\includegraphics[scale=.80]{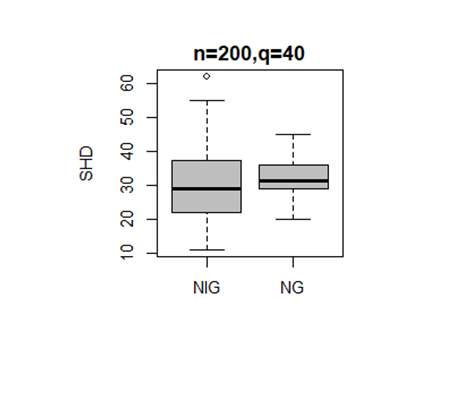}&\hspace{-.097cm}\includegraphics[scale=.80]{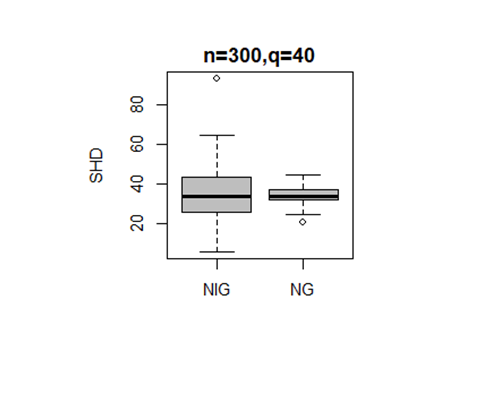}\vspace{0.5cm}
	\end{tabular}\vspace{-2.5cm}
	\label{figure1}
\end{figure}
\begin{figure}[!ht]\hspace{-1.3cm}
	\begin{tabular}{cc}
		\includegraphics[scale=0.95]{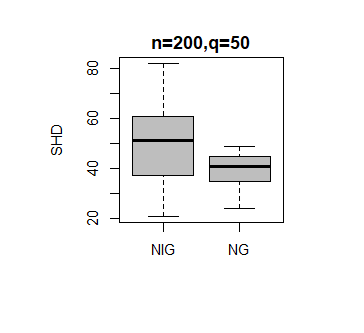}&\hspace{-.097cm}\includegraphics[scale=0.85]{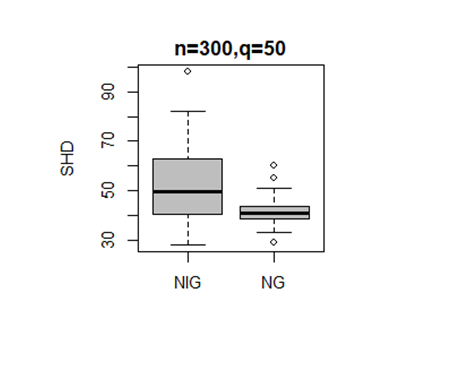}\vspace{0.5cm}
	\end{tabular}\vspace{-1.8cm}
	\caption{Boxplots of the SHD between the estimated DAG and the true DAG.}
	\label{figure2}
\end{figure}\\
Table 2 reflects that the SP, FNR, MISR, and AC measures for our model have improved when compared to the model based on the NIG prior. According to Figure \ref{figure2}, the structural Hamming distances (SHD) indicate that our model is superior since the amount of variations in SHD values for our model are less compared to the NIG model.
However, it's noteworthy that the SE and F1 values of the NG prior are lower when compared to the model NIG. This is attributed to the fact that our model has a very low TP, nearly approaching zero. Obtaining such results appears logical because we considered 
$\sfrac{3}{2q-2}=0.03$, that is, we considered the prior probability of edge inclusion in the model to be low, so the number of 1 values in the model decreases, causing the TP values of our model to shrink.\\
To demonstrate the impact of increasing $w$ on the results, similar to \cite{Castelletti6}, we set $w=0.2$. 
That is, we increase the probability of edge inclusion in the DAG. As indicated in Table 3, the outcome is denser DAGs compared to the previous configuration.
\begin{table}[!h]\label{table3}
	\small
	\begin{center}
		\begin{tabular}{|c|c|c|}
			\hline & NIG & NG \\
			\hline SE & 0.42 & 0.006 \\
			\hline F1 & 0.303 & 0.012\\
			\hline
		\end{tabular}
	\end{center}
	\caption{SE and F1 values for the NIG and NG prior with $q=40$ and $n=200$.}
\end{table}\\
In conclusion, the proposed method is more conservative in the inclusion of edges compared with the NIG prior model, as indicated by its lower sensitivity and F1-score. To elaborate more, in what follows, we show that it is possible to achive larger values than those of Table 3 for the SE and F1 in model. 
To this end, knowing that graph structures are not often used for DAGs, we computed these two measures for three well-known structures.
Table 4 illustrates how one achieves larger SE and F1 values using the proposed prior structure.
\begin{table}[!h]\label{table4}
	\small
	\begin{center}
		\begin{tabular}{|c|c|c|}
			\hline Graph structure & SE & F1 \\
			\hline Random & 0.13 & 0.21 \\
			\hline Hub & 0.21 & 0.11\\
			\hline Band & 0.16 & 0.22 \\
			\hline
		\end{tabular}
	\end{center}
	\caption{The SE and F1 values, for the NG prior with $q=40$ and $n=200$.}
\end{table}\\
Finally, Figure \ref{figure5} represents the computational time (CT) (averaged over 10 replicates) per iteration, as a function of $q\in \lbrace 5,10,20,50,100 \rbrace$ for $n=500$ (left panel), and as a function of $n\in\lbrace 50,100,200,300,500 \rbrace$ for $q=50$ (right panel). Surprisingly, the CT of our method increases with the number of variables and sample size, but it remains generally lower than that of the NIG model. Reflecting the Bessel function in the posterior and marginal models does not add computational delay or complexity.
\clearpage
\begin{figure}[!h]\hspace{-1.2cm}
	\begin{tabular}{cc}
		\includegraphics[scale=.38]{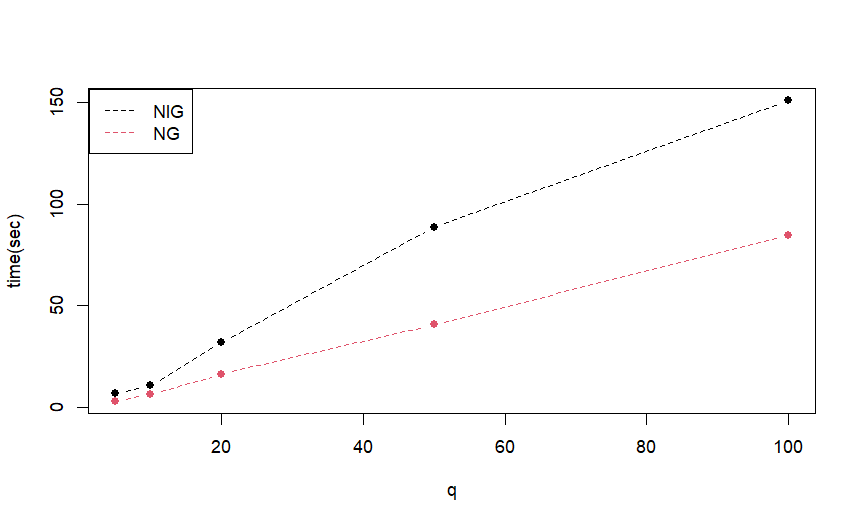}&\hspace{-.097cm}\includegraphics[scale=.38]{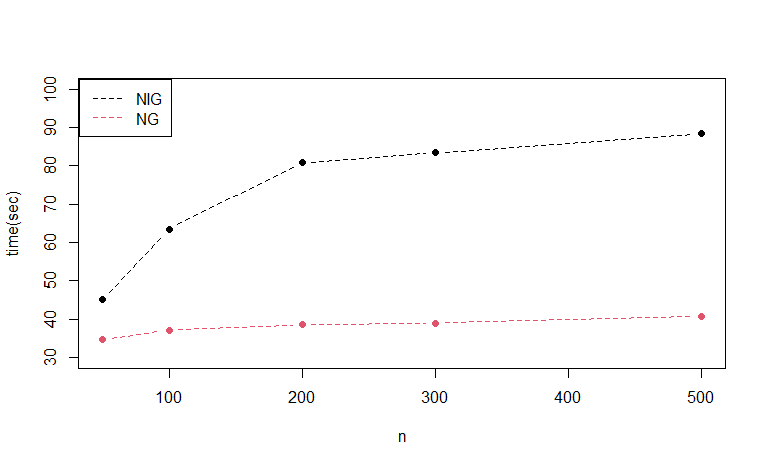}\vspace{0.5cm}
	\end{tabular}\vspace{-1cm}
	\caption{Computational time (in seconds) per iteration, as a function of the number of variables $q$ for fixed $n=500$ (left plot) and as a function of the sample size $n$ for fixed $q=50$ (right plot), averaged over 10 replicates.}
	\label{figure5}
\end{figure}
\subsection{Leukemia data analysis} 
In this subsection, we consider a biological dataset concerning patients affected by Acute Myeloid Leukemia (AML). AML represents an aggressive hematological cancer characterized by uncontrolled proliferation of hematopoietic stem cells in the bone marrow. 
AML responds very poorly to chemoterapeutic treatments, with a 5-year overall survival rate of about 25\%; the development of new
targeted treatments therefore represents a key strategy to improve patients' prospects.

\begin{figure}[!h]\hspace{0.3cm}
	\begin{tabular}{cc}
		\includegraphics[scale=0.8]{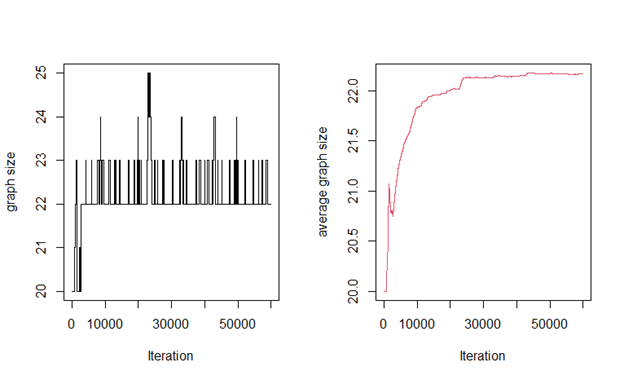}
	\end{tabular}\vspace{-0.6cm}
	\caption{Trace plot of the number of edges (graph size) in the DAGs visited
		by the MCMC at each iteration (left panel) and trace plot of the average number of edges in the
		DAGs computed up to iteration $s$, $s = 1, . . . , 60000$ (right panel) for AML data.}
	\label{figure6}
\end{figure}

\begin{figure}[!h]\hspace{0.3cm}
	\begin{tabular}{cc}
		\includegraphics[scale=0.7]{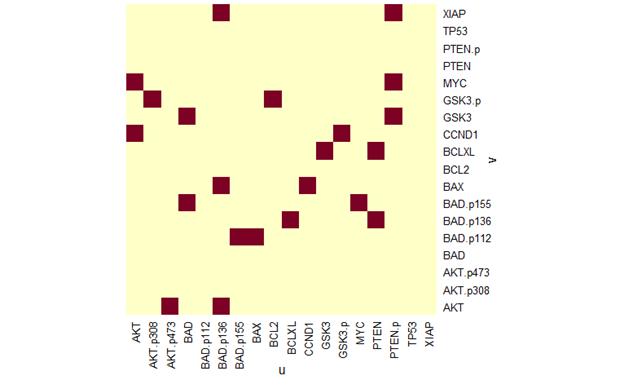}
	\end{tabular}\vspace{-0.6cm}
	\caption{Leukemia data. Heat map with posterior probabilities of edge inclusion, for each
		directed edge $u\rightarrow v$.}
	\label{figure7}
\end{figure}

\begin{figure}[!h]\hspace{-2.3cm}
	\begin{tabular}{cc}
		\includegraphics[scale=0.99]{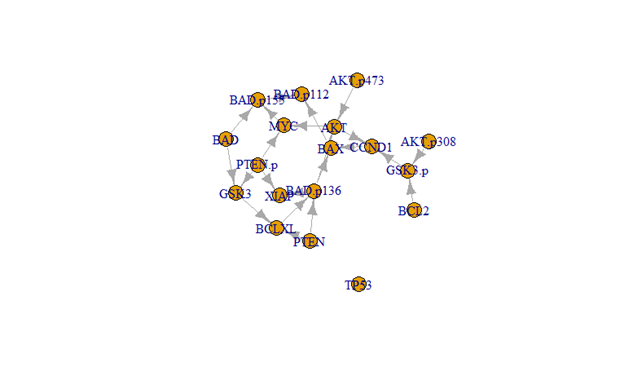}
	\end{tabular}\vspace{-2.5cm}
	\caption{The MPM estimate for AML dataset.}
	\label{figure8}
\end{figure}

\begin{figure}[!h]\hspace{-2.1cm}
	\begin{tabular}{cc}
		\includegraphics[scale=0.99]{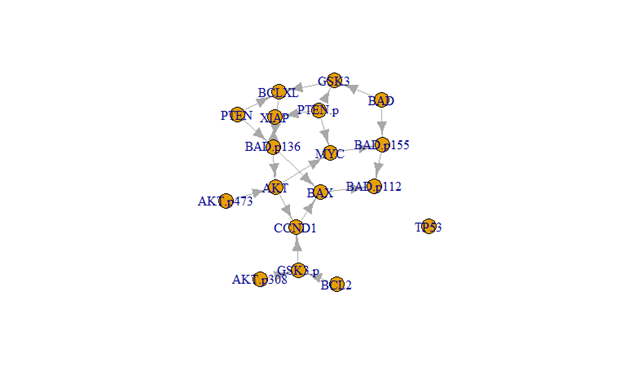}
	\end{tabular}\vspace{-2.3cm}
	\caption{The MAP estimate for AML dataset.}
	\label{figure9}
\end{figure}

The dataset analyzed here includes protein and phosphoproteins levels for 256 newly diagnosed AML patients and is provided as a supplement to \cite{Kornblau}. 
In addition, subjects are classified according to the French-American-British (FAB) system into several different AML subtypes.
The dataset contains the levels of $q=18$ proteins and phosphoproteins involved in apoptosis and cell cycle regulation according to the KEGG database \cite{Kanehisa}.\\
This data is analyzed by \cite{Peterson} and \cite{Castelletti4} from a multiple graphical model perspective to estimate group-specific dependence structures based on undirected and directed graphs respectively. Additionally \cite{Castelletti5} utilize this dataset for Bayesian graphical modeling to examine heterogeneous causal effects.\\
We focus on subtype M2 and $n=68$ samples from R package BCDAG. We run the MCMC algorithm with fixed parameters $S=60000$, $B=5000$, $a=q$, $U=\tfrac{1}{n}\boldsymbol{I}_{q}$, and $w=0.5$.\\
The trace plot of the number of edges in the graphs visited by the MCMC at each iteration $s = 1,..., S$ is shown in the left-side panel of Figure \ref{figure6}, and the trace plot demonstrating the average number of edges in the DAGs visited by the MCMC up to iteration $s$, for $s = 1,..., S$, is shown in the right-side panel.\\
The apparent absence of trends in the trace plot and the curve stabilization around an average value both suggest a good degree of MCMC mixing and convergence to the target distribution. Also Figure \ref{figure6} (right panel) shows that the average number of edges in the DAGs is around 22.
\\
Now we use the MCMC output to estimate the posterior probability of inclusion for each directed edge $u\rightarrow v$, that we report in the heatmap shown in Figure \ref{figure7}.
\newpage
Single DAG estimates that summarize the MCMC output can also be retrieved. For instance consider the maximum a posteriori (MAP) DAG estimate, which corresponds to the DAG with the highest estimated posterior probability. Alternatively, we can create the median probability model (MPM) by including any edges $u\rightarrow v$ with an estimated probability of inclusion greater than 0.5.\\
Figures \ref{figure8} and \ref{figure9} depict the MPM and MAP DAG estimates, respectively.
\newpage
The Maximum A Posteriori (MAP) model estimate is defined as the DAG visited by the MCMC with the highest associated posterior probability. The posterior probability of each DAG is estimated as the frequency of visits of the DAG in the MCMC chain. The MAP estimate is represented by its $q\times q$ adjacency matrix, with $\left(u,v\right)$ element equal to one whenever the MAP contains $u\rightarrow v$, zero otherwise.\\
The posterior probabilities corresponding to the NIG and NG models are 0.0016 and 0.769, respectively. It is evident that the posterior probability for our NG model is much higher compared to the NIG model.
\section{Conclusion}
This study presented a technique for learning the relationships between variables by utilizing Bayesian Directed Acyclic Graphs (DAGs) with the non-conjugate normal-gamma (NG) prior. Simulation experiments indicated that the suggested model is conservative in terms of edge inclusion. This means that it tends to construct sparse DAGs when the chance of including edges is low. We classified it as an adaptive Bayesian structure learning approach, where edges are included in the DAG if the inclusion probability adjusts to the true model. Our simulation studies have shown that our model produces a maximum a posteriori DAG (MAP DAG) with a higher posterior probability. Furthermore, our model outperforms the conjugate normal-inverse gamma (NIG) prior in terms of graphical metrics for learning the structure of Gaussian DAGs. In addition, our Bayesian technique exhibits reduced computational time for learning DAGs compared to the NIG prior. Our concept has potential applications in the fields of causal discovery and estimation. 



\section*{Appendix}
\label{sec:sample:appendix}
In this section, we provide the proof of Theorem \ref{th2}. To accomplish this, we first give some technical tools, as follows.\\
Let
$\boldsymbol{x},\boldsymbol{b}\in\mathbb{R}^{d}$
be two vectors, 
$\boldsymbol{M}\in\mathbb{R}^{d\times d}$
a symmetric and invertible matrix.\\
Then it can be seen easily that
\begin{equation*}
	\boldsymbol{x}^{T}\boldsymbol{M}\boldsymbol{x}-2\boldsymbol{b}^{T}\boldsymbol{x}=\left(\boldsymbol{x}-\boldsymbol{M}^{-1}\boldsymbol{b}\right)^{T}\boldsymbol{M}\left(\boldsymbol{x}-\boldsymbol{M}^{-1}\boldsymbol{b}\right)-\boldsymbol{b}^{T}\boldsymbol{M}^{-1}\boldsymbol{b}.\qquad\qquad\left(A1\right)
\end{equation*}
Let 
$K_{a}\left(.\right)$
be the Bessel function of the third kind (\cite{Gradshteyn}). Then for 
${\rm Re}(\alpha)>0$, ${\rm Re}(\beta)>0$, (where {\rm Re}(.) denotes the real part of a number)
\begin{equation*}
	\int_{0}^{\infty}u^{a-1}\exp\left(-\dfrac{\alpha}{u}-\beta u\right)du=2K_{a}\left(2\sqrt{\alpha\beta}\right)\left(\dfrac{\alpha}{\beta}\right)^{\tfrac{a}{2}}.\qquad\qquad\left(A2\right)
\end{equation*}
\begin{proof}[Proof of Theorem 1]
	By definition, using \eqref{eq4} and \eqref{eq7}, we have  
	\begin{align*}
		\hspace{-2cm}		m\left(\boldsymbol{X}_{j}\mid\boldsymbol{X}_{pa_{\mathcal{D}}\left(j\right)},\mathcal{D}\right)&=\int_{0}^{\infty}\int_{\mathbb{R}^{\lvert pa_{\mathcal{D}}\left(j\right)\rvert}}p\left(\boldsymbol{D}_{jj},\boldsymbol{L}_{\prec j ]}\mid\mathcal{D}\right)f\left(\boldsymbol{X}_{j}\mid\boldsymbol{X}_{pa_{\mathcal{D}}\left(j\right)},\boldsymbol{D}_{jj},\boldsymbol{L}_{\prec j ]},\mathcal{D}\right)d\boldsymbol{L}_{\prec j ]}d\boldsymbol{D}_{jj}\\
		&=\int_{0}^{\infty}\int_{\mathbb{R}^{\lvert pa_{\mathcal{D}}\left(j\right)\rvert}}p\left(\boldsymbol{L}_{\prec j ]}\mid\boldsymbol{D}_{jj},\mathcal{D}\right)p\left(\boldsymbol{D}_{jj}\mid\mathcal{D}\right)f\left(\boldsymbol{X}_{j}\mid\boldsymbol{X}_{pa_{\mathcal{D}}\left(j\right)},\boldsymbol{D}_{jj},\boldsymbol{L}_{\prec j ]},\mathcal{D}\right)d\boldsymbol{L}_{\prec j ]}d\boldsymbol{D}_{jj}\\
		&=\int_{0}^{\infty}\int_{\mathbb{R}^{\lvert pa_{\mathcal{D}}\left(j\right)\rvert}}\left(2\pi\right)^{-\tfrac{n}{2}}\big\lvert\boldsymbol{D}_{jj}\boldsymbol{I}_{n}\big\rvert^{-\tfrac{1}{2}}\\
		&\times\exp\biggr\lbrace-\dfrac{1}{2}\left(\boldsymbol{X}_{j}+\boldsymbol{X}_{pa_{\mathcal{D}}\left(j\right)}\boldsymbol{L}_{\prec j]}\right)^{T}\left(\boldsymbol{D}_{jj}\boldsymbol{I}_{n}\right)^{-1}\left(\boldsymbol{X}_{j}+\boldsymbol{X}_{pa_{\mathcal{D}}\left(j\right)}\boldsymbol{L}_{\prec j]}\right)\biggr\rbrace\\
		&\times\dfrac{\bigg(\dfrac{1}{2}\boldsymbol{U}_{jj\mid pa_{\mathcal{D}}\left(j\right)}\bigg)^{\tfrac{1}{2}\alpha_{j}^{\mathcal{D}}}}{\Gamma\left(\dfrac{1}{2}\alpha_{j}^{\mathcal{D}}\right)}\times\boldsymbol{D}_{jj}^{\tfrac{1}{2}\alpha_{j}^{\mathcal{D}}-1}\exp\biggr\lbrace-\dfrac{\boldsymbol{U}_{jj\mid pa_{\mathcal{D}}\left(j\right)}}{2}\boldsymbol{D}_{jj}\biggr\rbrace\times\left(2\pi\right)^{-\tfrac{\lvert pa_{\mathcal{D}}\left(j\right)\rvert}{2}}\biggr\vert\boldsymbol{D}_{jj}\boldsymbol{U}_{\prec j \succ}^{-1}\biggr\rvert^{-\tfrac{1}{2}}\\
		&\times\exp\biggr\lbrace-\dfrac{1}{2}\left(\boldsymbol{L}_{\prec j]}+\boldsymbol{U}_{\prec j \succ}^{-1}\boldsymbol{U}_{\prec j]}\right)^{T}\left(\boldsymbol{D}_{jj}\boldsymbol{U}_{\prec j \succ}^{-1}\right)^{-1}\left(\boldsymbol{L}_{\prec j]}+\boldsymbol{U}_{\prec j \succ}^{-1}\boldsymbol{U}_{\prec j]}\right)\biggr\rbrace d\boldsymbol{L}_{\prec j ]}d\boldsymbol{D}_{jj}\\
		&=\int_{0}^{\infty}\int_{\mathbb{R}^{\lvert pa_{\mathcal{D}}\left(j\right)\rvert}}\left(2\pi\right)^{-\tfrac{n+\lvert pa_{\mathcal{D}}\left(j\right)\rvert}{2}}\times\boldsymbol{D}_{jj}^{\tfrac{1}{2}\alpha_{j}^{\mathcal{D}}-\tfrac{\lvert pa_{\mathcal{D}}\left(j\right)\rvert}{2}-\tfrac{n}{2}-1}\times\big\lvert\boldsymbol{U}_{\prec j \succ}\big\rvert^{\tfrac{1}{2}}\\
		&\times\dfrac{\bigg(\dfrac{1}{2}\boldsymbol{U}_{jj\mid pa_{\mathcal{D}}\left(j\right)}\bigg)^{\tfrac{1}{2}\alpha_{j}^{\mathcal{D}}}}{\Gamma\left(\dfrac{1}{2}\alpha_{j}^{\mathcal{D}}\right)}
		\times\exp\biggr\lbrace-\dfrac{\boldsymbol{U}_{jj\mid pa_{\mathcal{D}}\left(j\right)}}{2}\boldsymbol{D}_{jj}\biggr\rbrace\\
		&\times\exp\biggr\lbrace-\dfrac{1}{2}\left(\boldsymbol{X}_{j}+\boldsymbol{X}_{pa_{\mathcal{D}}\left(j\right)}\boldsymbol{L}_{\prec j]}\right)^{T}\left(\boldsymbol{D}_{jj}\boldsymbol{I}_{n}\right)^{-1}\left(\boldsymbol{X}_{j}+\boldsymbol{X}_{pa_{\mathcal{D}}\left(j\right)}\boldsymbol{L}_{\prec j]}\right)\\
		&-\dfrac{1}{2}\left(\boldsymbol{L}_{\prec j]}+\boldsymbol{U}_{\prec j \succ}^{-1}\boldsymbol{U}_{\prec j]}\right)^{T}\left(\boldsymbol{D}_{jj}\boldsymbol{U}_{\prec j \succ}^{-1}\right)^{-1}\left(\boldsymbol{L}_{\prec j]}+\boldsymbol{U}_{\prec j \succ}^{-1}\boldsymbol{U}_{\prec j]}\right)\biggr\rbrace d\boldsymbol{L}_{\prec j ]}d\boldsymbol{D}_{jj}
	\end{align*}
Let
$\boldsymbol{M}=\widetilde{\boldsymbol{U}}_{\prec j \succ}=\boldsymbol{U}_{\prec j \succ}+\boldsymbol{X}_{pa_{\mathcal{D}}\left(j\right)}^{T}\boldsymbol{I}_{n}\boldsymbol{X}_{pa_{\mathcal{D}}\left(j\right)}$, and $\boldsymbol{b}=-\boldsymbol{X}_{pa_{\mathcal{D}}\left(j\right)}^{T}\boldsymbol{I}_{n}\boldsymbol{X}_{j}-\boldsymbol{U}_{\prec j ]}$,\\
by using $\left(A1\right)$ we have
\begin{align*}
	\hspace{-1.5cm}
	&\left(\boldsymbol{L}_{\prec j]}+\boldsymbol{U}_{\prec j \succ}^{-1}\boldsymbol{U}_{\prec j]}\right)^{T}\boldsymbol{U}_{\prec j \succ}\left(\boldsymbol{L}_{\prec j]}+\boldsymbol{U}_{\prec j \succ}^{-1}\boldsymbol{U}_{\prec j]}\right)
	+\left(\boldsymbol{X}_{j}+\boldsymbol{X}_{pa_{\mathcal{D}}\left(j\right)}\boldsymbol{L}_{\prec j]}\right)^{T}\boldsymbol{I}_{n}\left(\boldsymbol{X}_{j}+\boldsymbol{X}_{pa_{\mathcal{D}}\left(j\right)}\boldsymbol{L}_{\prec j]}\right)\\
	&=\boldsymbol{L}_{\prec j]}^{T}\left(\boldsymbol{U}_{\prec j \succ}+\boldsymbol{X}_{pa_{\mathcal{D}}\left(j\right)}^{T}\boldsymbol{I}_{n}\boldsymbol{X}_{pa_{\mathcal{D}}\left(j\right)}\right)\boldsymbol{L}_{\prec j]}+2\left(\boldsymbol{X}_{j}^{T}\boldsymbol{I}_{n}\boldsymbol{X}_{pa_{\mathcal{D}}\left(j\right)}+\boldsymbol{U}_{\prec j ]}^{T}\right)\boldsymbol{L}_{\prec j]}\\
	&+\boldsymbol{U}_{\prec j]}^{T}\boldsymbol{U}_{\prec j \succ}^{-1}\boldsymbol{U}_{\prec j]}+\boldsymbol{X}_{j}^{T}\boldsymbol{I}_{n}\boldsymbol{X}_{j}\\
	&=\left(\boldsymbol{L}_{\prec j]}-\boldsymbol{M}^{-1}\boldsymbol{b}\right)^{T}\boldsymbol{M}\left(\boldsymbol{L}_{\prec j]}-\boldsymbol{M}^{-1}\boldsymbol{b}\right)-\boldsymbol{b}^{T}\boldsymbol{M}^{-1}\boldsymbol{b}+\boldsymbol{U}_{\prec j]}^{T}\boldsymbol{U}_{\prec j \succ}^{-1}\boldsymbol{U}_{\prec j]}+\boldsymbol{X}_{j}^{T}\boldsymbol{I}_{n}\boldsymbol{X}_{j}
\end{align*}
Therefore, we obtain
\begin{eqnarray*}
m\left(\boldsymbol{X}_{j}\mid\boldsymbol{X}_{pa_{\mathcal{D}}\left(j\right)},\mathcal{D}\right)
	&=&\int_{0}^{\infty}\int_{\mathbb{R}^{\lvert pa_{\mathcal{D}}\left(j\right)\rvert}}\left(2\pi\right)^{-\tfrac{n+\lvert pa_{\mathcal{D}}\left(j\right)\rvert}{2}}\times\boldsymbol{D}_{jj}^{\tfrac{1}{2}\alpha_{j}^{\mathcal{D}}-\tfrac{\lvert pa_{\mathcal{D}}\left(j\right)\rvert}{2}-\tfrac{n}{2}-1}\cr
	&&\times\big\lvert\boldsymbol{U}_{\prec j \succ}\big\rvert^{\tfrac{1}{2}}
	\times\dfrac{\bigg(\dfrac{1}{2}\boldsymbol{U}_{jj\mid pa_{\mathcal{D}}\left(j\right)}\bigg)^{\tfrac{1}{2}\alpha_{j}^{\mathcal{D}}}}{\Gamma\left(\dfrac{1}{2}\alpha_{j}^{\mathcal{D}}\right)}\cr
	&&\times\exp\biggr\lbrace-\dfrac{\boldsymbol{U}_{jj\mid pa_{\mathcal{D}}\left(j\right)}}{2}\boldsymbol{D}_{jj}-\dfrac{1}{2\boldsymbol{D}}_{jj}\biggr\lbrack\left(\boldsymbol{L}_{\prec j]}-\boldsymbol{M}^{-1}\boldsymbol{b}\right)^{T}\boldsymbol{M}\left(\boldsymbol{L}_{\prec j]}-\boldsymbol{M}^{-1}\boldsymbol{b}\right)\cr
	&&-\boldsymbol{b}^{T}\boldsymbol{M}^{-1}\boldsymbol{b}+\boldsymbol{U}_{\prec j]}^{T}\boldsymbol{U}_{\prec j \succ}^{-1}\boldsymbol{U}_{\prec j]}+\boldsymbol{X}_{j}^{T}\boldsymbol{I}_{n}\boldsymbol{X}_{j}\biggr\rbrack\biggr\rbrace d\boldsymbol{L}_{\prec j ]}d\boldsymbol{D}_{jj}
\end{eqnarray*}
Integrating with respect to
$\boldsymbol{L}_{\prec j ]}$ and $\boldsymbol{D}_{jj}$ respectively, using $\left(A2\right)$ gives the required result.\\
\end{proof}

\section*{Author Contribution}
Conceptualization, S.N. and M.A; Methodology, S.N., M.A. and A.S.; Software, S.N. and M.A.; Supervision, M.A. and A.S.; Formal analysis, S.N., M.A. and A.S.; Writing—original draft, S.N.; Writing—review and editing, S.N., M.A. and A.S. 
\section*{Ethical Statement}
The authors declare no conflict of interest.
\section*{Data Availability Statement}
Data is available in the numerical analysis section. 
\section*{Funding Statement}
Mohammad Arashi’s work is based on the research supported in part by the Iran National Science Foundation (INSF) grant No. 4015320.

\section*{ORCID}
\noindent Mohammad Arashi (0000-0002-5881-9241)\\
Abdolnasser Sadeghkhani (0000-0002-3074-0443)

\bibliographystyle{elsarticle-num-names} 


\end{document}